# Cord-blood vitamin D level and night sleep duration in preschoolers in the EDEN mother-child birth cohort

Chu Yan Yong[1,2], Eve Reynaud[1,2], Anne Forhan[1,2], Patricia Dargent-Molina[1,2], Barbara Heude[1,2], Marie-Aline Charles[1,2], Sabine Plancoulaine[1,2]; on behalf of the EDEN study group.

**Affiliations:**

[1] INSERM, UMR1153, Epidemiology and Statistics Sorbonne Paris Cité Research Center (CRESS), early ORigins of Child Health And Development Team (ORCHAD), Villejuif, F-94807 France;

[2] Univ Paris-Descartes, UMRS 1153, Paris, France;

**Address correspondence to:**

Sabine Plancoulaine, INSERM, UMR1153, Epidemiology and Statistics Sorbonne Paris Cité Research Center (CRESS), early ORigins of Child Health And Development Team (ORCHAD), 16 Av Paul Vaillant Couturier, 94 807 Villejuif Cedex, FRANCE,

Email: sabine.plancoulaine@inserm.fr

Phone: + 33 1 45 59 51 09.

# ABSTRACT

**Objective:** 25-hydroxyvitamin D (25OHD) deficiency has been associated with sleep disorders in adults. Only three cross-sectional studies were performed in children and showed an association between 25OHD deficiency and both obstructive sleep apnea syndrome and primary snoring. No longitudinal study has been performed in children from the general population. We analyzed the association between cord-blood vitamin D level at birth and night-sleep duration trajectories for children between 2 and 5-6 years old in a non-clinical cohort.

**Method:** We included 264 children from the French EDEN mother-child birth-cohort with both cord-blood 25OHD level determined by radio-immunoassay at birth, and night-sleep trajectories for children between 2 and 5-6 years old obtained by the group-based trajectory modeling method. Associations between 25OHD and sleep trajectories were assessed by multinomial logistic regression adjusted for maternal and child characteristics.

**Results:** The trajectories short sleep (<10h30/night), medium-low sleep (10h30-11h00/night), medium-high sleep (≈11h30/night), long sleep (≥11h30/night) and changing sleep (decreased from ≥11h30 to 10h30-11h00/night) represented 5%, 46%, 37%, 4% and 8% of the children, respectively. The mean 25OHD level was 19 ng/ml (SD=11, range 3 to 63). It was 12 (SD=7), 20 (SD=11), 19 (SD=10), 14 (SD=7) and 16 (SD=8) ng/ml for children with short, medium-low, medium-high, long and changing sleep trajectories, respectively. On adjusted analysis, for each 1-ng/ml decrease in 25OHD level, the odds of belonging to the short sleep versus medium-high sleep trajectory was increased (odds ratio =1.12, 95% confidence interval [1.01-1.25]). We found no other significant association between 25OHD level and other trajectories.

**Conclusion:** Low 25OHD level at birth may be associated with increased probability of being a persistent short sleeper in preschool years. These results need confirmation.

Keywords: pediatric sleep, vitamin D, epidemiology, cohort,

## 1. Introduction

Vitamin D, a steroid hormone, is involved in bone metabolism promoting digestive calcium absorption, apoptosis and angiogenesis, decreasing cell proliferation. It has an immunomodulatory, anti-infectious, anti-inflammatory and anti-tumor role. Vitamin D deficiency has been associated with many pathological conditions including osteoporosis, microbial infections, cardiovascular diseases, cancers, autoimmune diseases, asthma and allergy (reviewed in [1]). It is common in the general population (up to 80% of European adolescents) [2,3]. The known risk factors for vitamin D deficiency in adults and children in the general population are pigmented skin, obesity, self-limitation of solar exposure and use of sunscreens, and poor dietary intake of vitamin D (e.g., fatty fish, egg, milk) [4,5]. There is a seasonal variation in vitamin D levels [6], with higher levels in summer and lower levels in winter.

Cross-sectional studies have recently suggested a role for vitamin D and its metabolism in sleep [7,8], in particular in the development of symptoms such as obstructive sleep apnea [9], diurnal somnolence [10], and restless leg syndrome [11] in adults. In older people, vitamin D deficiency was found associated with poor sleep (short duration or low efficiency) [12] that was improved by supplementation [13]. In children, only three cross-sectional studies have been published and showed associations between vitamin D deficiency and obstructive sleep apnea syndrome and primary snoring [14–16].

Here we aimed to analyze the association between cord-blood vitamin D level at birth and night-sleep duration trajectories in children between 2 and 5-6 years old in a non-clinical cohort.

## 2. Material and methods
### 2.1. Study population

The EDEN study aims at investigating the pre- and post-natal determinants of child health and development. Details of the EDEN study protocol have been previously published [17]. Briefly, pregnant women under 24 weeks of amenorrhea were recruited between 2003 and 2006 in the university hospitals of Poitiers and Nancy. Those under 18 years, unable to give informed consent, functionally illiterate in French, with a history of diabetes, planning on changing address or without social security coverage were excluded from the cohort. Women with multiple pregnancies were also excluded. A total of 1899 children were enlisted at birth. Written informed consent was obtained twice from parents: at enrolment and after the child's birth. The study was approved by an ethics research committee and by the national data protection authority.

### 2.2. Measures and participant characteristics

#### 2.2.1. Cord-blood vitamin D measurement

Cord-blood samples were collected immediately after birth (vaginal delivery) or after extraction of the fetus via uterine incision (elective cesarean section) and were centrifuged within 24 h of collection. The serum was separated and stored at -80°C. Serum 25-hydroxyvitamin D (25OHD), representative of overall vitamin D stored in the body, was measured by immunochemiluminescent immunoassay performed on the LIAISON platform (DiaSorin, Sallugia, Italy). The intra- and inter-assay coefficient of variation was < 10% whatever the measured level. This measure was performed for a subsample of 375 children from the EDEN cohort that correspond to infants who had quantitative ultrasonography measurements of bone status at age 1 year (i.e., infants examined from April 2006 onward) [18].

#### 2.2.2. Night-sleep duration trajectories in children between 2 and 5-6 years old

Night-sleep duration was collected at age 2, 3 and 5-6 years by using parental self-administered questionnaires and was calculated on the basis of the answers to the following questions: "Usually, at what time does your child go to bed?", "Usually, at what time does your child wake up?". Responses were recorded in hours and minutes (e.g., 10h30). "Group-based trajectory modeling" developed by Nagin et al. [19], implemented under SAS (PROC TRAJ) and data-driven, was used to identify night-sleep duration trajectories among 1205 children from the cohort whose parents had answered the questions regarding night-sleep durations for at least two of three age points. The method is based on the underlying hypothesis that within a population there are inherent groups that evolve according to different sleep patterns. The groups are not directly identifiable or pre-established by sets of characteristics but are statistically determined by each series of responses by using maximum likelihood.

Five night-sleep duration trajectories were established as previously reported [20] (Figure 1): short sleep (SS, <10h30/night, 4.9% of 1205 children), medium-low sleep (MLS, 10h30-11h00/night, 47.8%), medium-high sleep (MHS, about 11h30/night, 37.2%), long sleep (LS, ≥11h30/night, 4.5%) and changing sleep (CS, i.e., LS then MLS, 5.6%). Each child was assigned to the trajectory to which he/she belonged with the highest probability. Only children with both an assigned trajectory and cord-blood vitamn D measure were included in the current study.

### 2.2.3. Socio-demographic and health characteristics

Household socio-economic and demographic factors as well as maternal characteristics were collected at inclusion: maternity ward of recruitment (Nancy/Poitiers), household monthly income (<1500, 1500-3000 and ≥3000 euros, US dollar equivalent: <$1600, $1600-3250, > $3250), maternal education level (< high-school, high school diploma to 2-year university degree, >2-year university degree), and maternal age at delivery. Body mass index (BMI) before pregnancy was calculated by using reported height and weight. Child's sex and season of birth was collected from maternity medical charts. Because of French regulations, ethnic origin was not collected. However, we collected information on geographic origin of parents and grandparents. Children were considered of European origin if both parents and maternal grandparents were born in a European country.

### 2.3. Statistical analysis

A total of 264 children presented available data for both 25OHD measure and sleep trajectory and were included in the present analysis. They were compared to non-included children for maternal and child characteristics by chi-square and Student *t* test. The associations between socio-demographic and health characteristics and night-sleep duration trajectories and between cord blood vitamin D level and night-sleep trajectories were assessed by multiple multinomial logistic regression (SAS 9.3, SAS Institute Inc, Cary, NC, USA). Multivariable models estimated odds ratios (ORs) and 95% confidence intervals (CIs) associated with a 1-unit decrease in 25OHD level. Confounding factors were identified from the literature and selected by using the Directed Acyclic Graphs method (www.dagitty.net) [21]. The resulting model adjusted for recruitment centre, maternal education, familial income, family geographical origin, pre-pregnancy maternal BMI, maternal age at delivery, child's sex and season of birth.

## 3. Results

Compared to non-included children, for the 264 children in the study, mothers were older (30 vs 29 years, p=0.003) and more educated (41% vs 30% >2-year university degree, p<0.0001) and had higher incomes (32% vs 26% with income >3000 euros, p<0.0002). Included children were frequently boys (60% vs 51%, p=0.02), born in spring (43% vs 28%, p<0.0001) and had higher mean 25OHD level (19 vs 15 ng/ml, p=0.003). Distribution of night-sleep duration trajectories did not differ between included and excluded children (p=0.47). Table 1 provides the characteristics of included children. The mean 25OHD level was 12 (SD=7), 20 (SD=11), 19 (SD=10), 16 (SD=8), and 14 (SD=7) ng/ml for the SS, MLS, MHS, CS and LS trajectories, respectively. Crude and adjusted ORs of belonging to a given sleep trajectory versus the MHS trajectory (reference) are presented in Figure 2. After adjustment, ORs remained stable and each 1-ng/ml decrease in 25OHD level was associated with 12% increased odds of belonging to the SS trajectory (vs the MHS trajectory). Only the estimated OR for belonging to the LS trajectory was modified by adjustment and increased from 1.06 [95% CI 0.98-1.14] to 1.10 [1.00-1.23] for each 1-ng/ml decrease in 25OHD level, remaining bordeline significant.

4. Discussion

This is the first longitudinal study exploring 25OHD level in newborns from the general population and its association with sleep duration during preschool age. On adjusted analysis, for each 1-ng/ml decrease in 25OHD level, the odds of belonging to the SS versus MHS trajectory was increased 12%.

We report generally low serum 25OHD level at birth in this sample of children, which suggests global vitamin D deficiency in French newborns and their mothers. The American Association of Pediatrics estimated that 25OHD level should be ≥20 ng/ml in infants and children [22], whereas the Endocrine Society recommends a 25OHD level >30 ng/ml [23]. In our study, only 41% and 16% of children presented such 25OHD levels at birth, respectively. The prevalence of 25OHD deficiency (<20 ng/ml) was lower than that reported in infant cord blood in the United States [24] and in a recent French study [25], which showed about two thirds of newborns with 25OHD level <20 ng/ml. This discrepancy may be due to the population selection at inclusion, which thus differed from the targeted population [17] and the follow-up as described, with older mothers, having higher incomes and education and higher 25OHD levels in included than excluded children.

Cord-blood 25OHD levels differed according to night-sleep trajectories. We observed a mean level of about the recommended level of 20 ng/ml for MHS and MLS trajectories for children between 2 and 5-6 years old; these two trajectories are nearest to the recommended sleep durations for children of this age range [26]. Decreased cord-blood 25OHD level was associated with an increased odds of belonging to the SS trajectory between 2 and 5-6 years old (i.e., persistent night-sleep duration <10h30 between 2 and 5-6 years old), which suggests an early effect of 25OHD level on sleep or sleep regulation. The association between low 25OHD serum level and short sleep duration was shown in cross-sectional studies in adults, especially in older people in several countries (United States, Korea and Brazil) [12,27–29]. However, no study on sleep duration was performed in children.

Physiological links have been observed between vitamin D and sleep, suggesting that vitamin D has direct effects on the initiation and maintenance of sleep [30]. Indeed, the vitamin D receptor is involved in brain development [31] and has been found in many cerebral regions, including those that regulate sleep [32–38]. Trials of vitamin D supplementation in adults with sleep troubles showed sleep amelioration, including increased sleep duration [35,39]. However, vitamin D deficiency occurring in early childhood during brain development may lead to persistent sleep troubles.

25OHD easily crosses the placenta barrier, with a strong correlation between cord blood and maternal serum values [40]. The vitamin D pool of the fetus and newborn depends on their mother's vitamin D status. Vitamin D supplementation during pregnancy should limit vitamin D deficiency in infants and may favor both brain development and healthy sleep in children. While already applied during pregnancy, supplementation seems insufficient and should be reinforced [25]. Increased child's sleep duration with vitamin D supplementation needs further exploration.

The strengths of this study are the general population sample and the longitudinal data for sleep duration in children. This study also presents some limitations. The attrition discussed above restricts the generalization of the results. However, included and excluded children did not differ by sleep trajectory distribution, and 25OHD level was measured in the context of another study objective, with blinding to sleep data. Thus, the bias should be minimal. However, because the studied sample size was greatly reduced, the sampling fluctuations (i.e., confidence intervals) were increased and results need to be replicated in a larger population.

## 5. Conclusion

In this first longitudinal study exploring the relation between 25OHD level at birth and sleep duration in preschool years, we suggest that low 25OHD level is associated with increased odds of children in a French birth cohort between 2 and 5-6 years old to be persistent short sleepers. These results need to be confirmed in larger sample of children from the general population.

**Acknowledgments**

Collaborators: We thank the EDEN mother-child cohort study group (I. Annesi-Maesano, J.Y Bernard, J. Botton, M.A. Charles, P. Dargent-Molina, B. de Lauzon-Guillain, P. Ducimetière, M. de Agostini, B. Foliguet, A. Forhan, X. Fritel, A. Germa, V. Goua, R. Hankard, B. Heude, M. Kaminski, B. Larroque†, N. Lelong, J. Lepeule, G. Magnin, L. Marchand, C. Nabet, F. Pierre, R. Slama, M.J. Saurel-Cubizolles, M. Schweitzer, O. Thiebaugeorges).

We thank all funding sources for the EDEN study: Foundation for medical research (FRM), National Agency for Research (ANR), National Institute for Research in Public health (IRESP: TGIR cohorte santé 2008 program), French Ministry of Health (DGS), French Ministry of Research, INSERM Bone and Joint Diseases National Research (PRO-A) and Human Nutrition National Research Programs, Paris–Sud University, Nestlé, French National Institute for Population Health Surveillance (InVS), French National Institute for Health Education (INPES), the European Union FP7 programs (FP7/2007-2013, HELIX, ESCAPE, ENRIECO,Medall projects), Diabetes National Research Program (in collaboration with the French Association of Diabetic Patients (AFD), French Agency for Environmental Health Safety (now ANSES), Mutuelle Générale de l'Education Nationale complementary health insurance (MGEN), French national agency for food security, French speaking association for the study of diabetes and metabolism (ALFEDIAM).

**Table 1**. Description of the studied population (N=264) from the EDEN mother-child cohort by night-sleep duration trajectories.

| | Total n (%) or mean (SD) | SS (N=14, 5%) n (%) or mean (SD) | MLS (N=121, 46%) n (%) or mean (SD) | MHS (N=98, 37%) n (%) or mean (SD) | CS (N=20, 8%) n (%) or mean (SD) | LS (N=11, 4%) n (%) or mean (SD) | p-value |
|---|---|---|---|---|---|---|---|
| **Maternal characteristics** | | | | | | | |
| Recruitment center (Poitiers) | 122 (46%) | 5 (36%) | 52 (43%) | 49 (50%) | 7 (35%) | 9 (82%) | 0.08 |
| Family income (€/month) | | | | | | | 0.82 |
| <1500 | 22 (8%) | 1 (7%) | 10 (8%) | 8 (8%) | 3 (15%) | 0 (0%) | |
| [1500-3000] | 157 (60%) | 10 (72%) | 74 (61%) | 54 (55%) | 13 (65%) | 6 (55%) | |
| >3000 | 85 (32%) | 3 (21%) | 37 (31%) | 36 (37%) | 4 (20%) | 5 (45%) | |
| Maternal education level | | | | | | | 0.41 |
| <High school | 50 (19%) | 4 (28%) | 20 (17%) | 19 (19%) | 6 (30%) | 1 (9%) | |
| High school to 2-year university degree | 106 (40%) | 5 (36%) | 52 (43%) | 35 (36%) | 6 (30%) | 8 (73%) | |
| >2 year university degree | 108 (41%) | 5 (36%) | 49 (40%) | 44 (45%) | 8 (40%) | 2 (18%) | |
| Maternal age at delivery (years) | 30 (5) | 30 (5) | 30 (4) | 31 (5) | 30 (5) | 30 (4) | 0.87 |
| Pre-pregnancy BMI | 23 (4) | 23 (3) | 23 (4) | 23 (4) | 22 (6) | 24 (5) | 0.93 |
| **Child characteristics** | | | | | | | |
| Sex (boy) | 156 (60%) | 12 (86%) | 78 (64%) | 53 (54%) | 8 (40%) | 5 (45%) | 0.05 |
| Birth season | | | | | | | 0.89 |
| Spring | 113 (43%) | 5 (36%) | 50 (41%) | 46 (47%) | 8 (40%) | 4 (37%) | |
| Summer | 65 (25%) | 1 (7%) | 34 (28%) | 23 (24%) | 4 (20%) | 3 (27%) | |
| Autumn | 53 (20%) | 5 (36%) | 22 (18%) | 18 (18%) | 5 (25%) | 3 (27%) | |
| Winter | 33 (12%) | 3 (21%) | 15 (13%) | 11 (11%) | 3 (15%) | 1 (9%) | |
| Cord-blood 25OHD level (ng/ml) | | 19 (11) | 12 (7) | 20 (11) | 19 (10) | 16 (8) | 14 (7) | 0.02 |
| | <10 | 59 (23%) | 6 (43%) | 23 (19%) | 22 (23%) | 6 (30%) | 2 (18%) | |
| | 10-20 | 98 (37%) | 6 (43%) | 39 (32%) | 39 (40%) | 7 (35%) | 7 (64%) | |
| | 21-29 | 64 (24%) | 2 (14%) | 33 (27%) | 21 (21%) | 6 (30%) | 2 (18%) | |
| | ≥30 | 43 (16%) | 0 (0%) | 26 (22%) | 16 (16%) | 1 (5%) | 0 (0%) | |

* SS = short sleep (<10h30/night), MLS = medium-low sleep (10h30-11h00/night), MHS = medium-high sleep (about 11h30/night), CS = changing sleep (i.e., LS then MLS) and LS= long sleep (≥11h30/night).

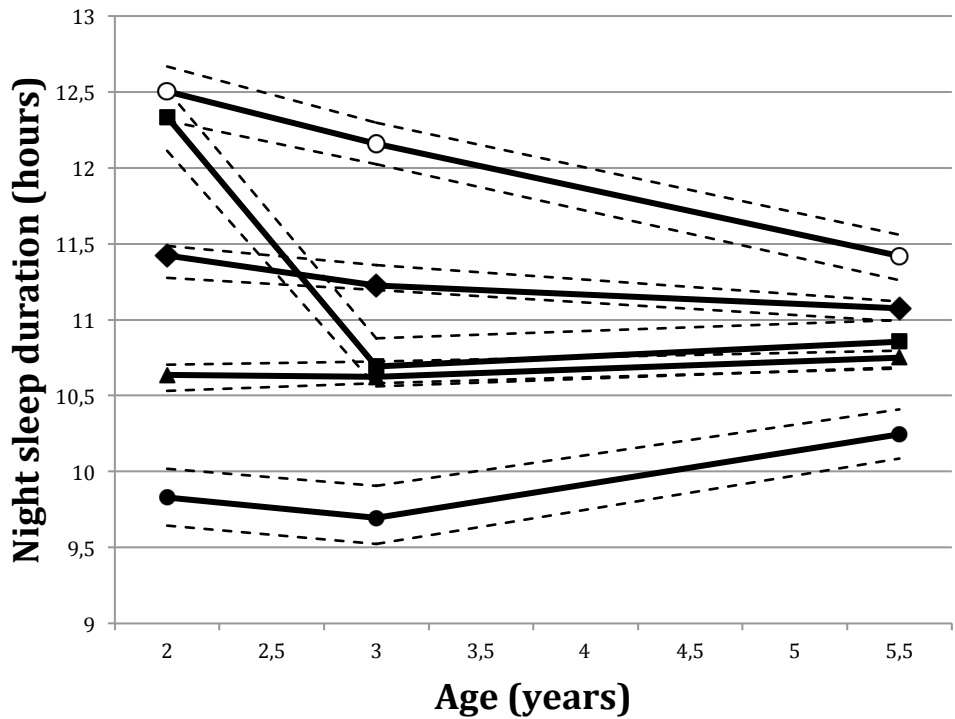

**Figure 1.** Night-sleep duration trajectories for EDEN preschool children (N=1205). Full lines represent mean sleep duration trajectories. Black circles = short sleep (SS, 4.9% of the children): triangles = medium-low sleep (MLS, 47.8% of the children); diamonds = medium-high sleep (MHS, 37.2% of the children), squares = changing sleep (CS, 5.6% of the children) and white circles = long sleep (LS, 4.5% of the children). Dashed lines represent the 95% confidence intervals for the trajectory estimations. Figure from Plancoulaine et al. [41].

**Figure 2.** Unadjusted (in grey) and adjusted (in black) odds ratios for 1-ng/ml decrease in 25OHD level by sleep duration trajectories. SS = short sleep (<10h30/night), MLS = medium-low sleep (10h30-11h00/night), MHS = medium-high sleep (about 11h30/night), CS = changing sleep (i.e., LS then MLS) and LS= long sleep (≥11h30/night). MHS is the reference trajectory.

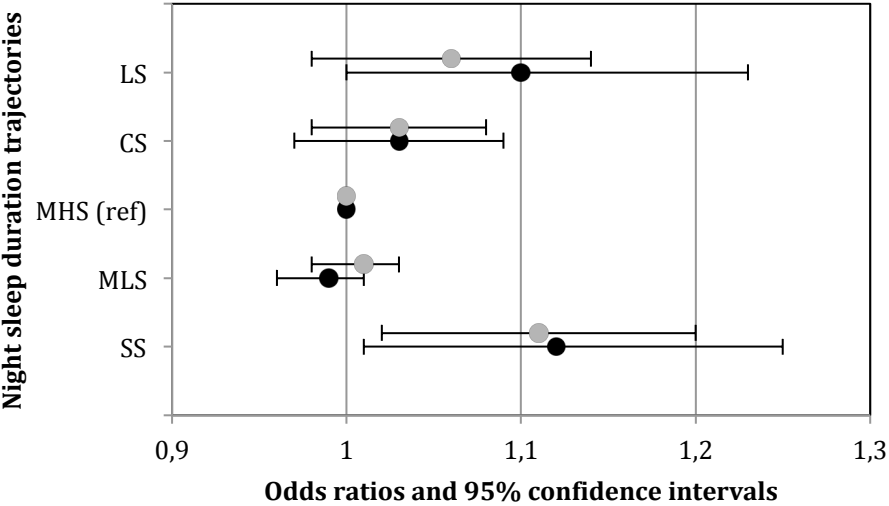